\documentclass[fleqn,10pt,final]{wlscirep}
\usepackage{amsmath,amssymb}
\usepackage{booktabs}
\usepackage{graphicx}
\usepackage{hyperref}
\usepackage[numbers]{natbib}
\usepackage{xcolor}
\usepackage{microtype}
\usepackage{array}
\usepackage{longtable}
\usepackage{subcaption}
\usepackage{float}
\usepackage{placeins}
\usepackage{multirow}
\graphicspath{{./}}

\title{Causal Inference of Blood Pressure Reduction and Coronary Heart Disease Risk in the Framingham Study}

\author[1,*]{Suchibrata Patra}
\affil[1]{St.Xavier's College (Autonomous), Kolkata}
\affil[*]{Correspondence: \href{mailto:official@suchibrata.in}{suchibratapatra2003@gmail.com}}


\begin{abstract}
Standard cardiovascular risk calculators, including the Framingham
Risk Score and the ACC/AHA Pooled Cohort Equations, estimate the
conditional probability $P(\mathrm{CHD} \mid \mathrm{SysBP}=s)$
rather than the interventional quantity
$P(\mathrm{CHD} \mid \mathrm{do}(\mathrm{SysBP}=s))$.
When confounding is present, this distinction has direct clinical
consequences: observational estimates may systematically overstate
the absolute benefit of antihypertensive treatment.
We applied Pearl's do-calculus to the Framingham Heart Study
Offspring Cohort ($n=4{,}240$; primary analysis on $3{,}776$
complete cases; 574 ten-year coronary heart disease (CHD) events).
A structurally corrected directed acyclic graph (DAG) was specified,
incorporating four biologically motivated corrections, and subjected
to conditional independence testing.
The average causal effect (ACE) of a 20~mmHg systolic blood pressure
(SysBP) reduction was estimated by g-computation with 1,500-iteration
bootstrap confidence intervals, corroborated by sex-stratified
propensity score matching (PSM) and inverse probability weighting (IPW).
Conditional average treatment effects (CATE) were estimated using R-Learner
and T-Learner metalearners with gradient-boosted nuisance models.
G-computation yielded an ACE of 3.40\% absolute risk reduction
(95\% bootstrap CI: 2.64\%--4.14\%), compared with a naive
observational estimate of 4.14\%, a relative overestimation of
approximately 21.8\% ($+35.7\%$ on the log-odds scale).
The E-value lower bound was 2.18.
Statistically significant heterogeneity in treatment effect was
detected across age strata (Kruskal--Wallis $p<0.001$) and diabetes
status (Mann--Whitney $p<0.001$); however, diabetic subgroup
estimates were unstable and underpowered ($n=109$), and no reliable
subgroup inference can be drawn without replication.
These findings suggest that observational cardiovascular risk tools
may overestimate the absolute benefit of blood pressure reduction,
with implications for clinical risk stratification and prescribing
thresholds.
\end{abstract}

\begin{document}

\flushbottom
\maketitle
\thispagestyle{empty}

% =======================================================
\section*{Introduction}
% =======================================================

Every major cardiovascular risk tool in routine clinical
use, the Framingham Risk Score~\cite{Wilson1998},
SCORE2~\cite{SCORE22021}, and the ACC/AHA Pooled Cohort
Equations~\cite{Goff2014}, estimates a conditional probability of
the form $P(\mathrm{CHD} \mid \mathrm{SysBP}=s,\, Z=z)$.
This quantity describes the observed frequency of coronary heart
disease (CHD) among individuals who happen to present with a given
blood pressure profile.
The quantity a prescribing clinician requires when deciding whether
to initiate antihypertensive therapy is categorically different:
\begin{equation}
  P\!\left(\mathrm{CHD} \mid \mathrm{do}(\mathrm{SysBP} \leftarrow s)\right),
  \label{eq:do}
\end{equation}
where Pearl's do-operator~\cite{Pearl2009} represents the act of
pharmacologically setting blood pressure to a target value,
regardless of the patient's underlying biological state.
Individuals who naturally present with $\mathrm{SysBP}=140$~mmHg
are, on average, older, more obese, and more metabolically
dysregulated than those presenting at 120~mmHg.
Antihypertensive therapy severs only the causal path from elevated
pressure to vascular events; it does not reverse these pre-existing
comorbidities.
Conflating the observational and interventional quantities therefore
produces a systematically inflated estimate of treatment benefit, a
gap with direct consequences for clinical decision-making and health
economic evaluation~\cite{Hernan2020}.

This prediction--intervention gap is formally resolvable using
structural causal models and the back-door adjustment
theorem~\cite{Pearl2009}.
Under this framework, the interventional distribution
(\ref{eq:do}) is identifiable from observational data provided that
a valid directed acyclic graph (DAG) can be specified and that the
positivity and conditional ignorability assumptions hold.
A growing methodological literature has addressed components of this
problem: marginal structural models have been applied to blood
pressure trajectories~\cite{Vansteelandt2012}, inverse probability
weighting (IPW) has been reviewed in cardiovascular
epidemiology~\cite{Mansournia2016}, and the distinction between
associational and interventional estimands has been extensively
theorised~\cite{Hernan2020,Naimi2017}.
Nevertheless, no prior study has delivered, within a single
reproducible pipeline applied to the Framingham Heart Study, a
biologically corrected and empirically evaluated causal DAG, multiple
corroborating causal-effect estimators, and an explicit
quantification of the bias incurred by standard observational
approaches.

The present study makes three contributions.
First, we quantify, for the first time in this cohort, the
magnitude by which the naive observational estimate of blood pressure
treatment benefit exceeds the causal estimate (approximately 21.8\%
in relative terms).
Second, we specify, correct, and empirically evaluate a structural
causal DAG, identifying four errors in prior formulations that
propagate into biased causal estimates.
Third, we implement an end-to-end causal inference pipeline, covering
DAG evaluation, g-computation, sex-stratified propensity score matching (PSM),
IPW, refutation testing, and semiparametric
treatment-effect heterogeneity estimation, within a single
reproducible framework, and discuss the clinical implications of the
resulting estimates.

% =======================================================
\section*{Results}
% =======================================================

\subsection*{Baseline characteristics}

The analytical cohort comprised 3,776 complete-case participants
(55.5\% female; mean age $49.6 \pm 8.6$ years; mean
systolic blood pressure (SysBP) $132.4 \pm 22.0$~mmHg; 2.7\% with diabetes
mellitus).
As shown in Table~\ref{tab:baseline}, participants who developed
CHD within ten years were on average five years older
($54.2$ vs.\ $48.7$ years; Mann--Whitney $p<0.001$), had 13.5~mmHg
higher SysBP ($143.8$ vs.\ $130.3$~mmHg;
$p<0.001$), and had substantially higher prevalences of diabetes
(6.3\% vs.\ 2.1\%; $\chi^2$, $p<0.001$) and prevalent hypertension
(50.7\% vs.\ 27.6\%; $p<0.001$).
These systematic differences illustrate the confounding structure
that motivates causal rather than observational estimation: patients
presenting with elevated SysBP are metabolically
and cardiovascularly more compromised across multiple dimensions,
independently of their blood pressure level.

\begin{table}[H]
\caption{\textbf{Baseline characteristics by 10-year CHD status.}
Data are mean $\pm$ SD (continuous variables) or $n$ (\%) (binary
variables). $p$-values from Mann--Whitney $U$ test (continuous) or
$\chi^2$ test (binary). The systematic confounding structure, with older
age, higher SysBP, higher body mass index (BMI), and greater comorbidity burden among
CHD cases, motivates causal rather than observational estimation.}
\label{tab:baseline}
\centering
\small
\begin{tabular}{lcccc}
\toprule
\textbf{Variable} & \textbf{Overall} & \textbf{No CHD} & \textbf{CHD} & \textbf{\textit{p}} \\
 & $(n=3{,}776)$ & $(n=3{,}202)$ & $(n=574)$ & \\
\midrule
Age (years)                & $49.6\pm 8.6$  & $48.7\pm 8.4$  & $54.2\pm 8.0$  & $<0.001$ \\
SysBP (mmHg)               & $132.4\pm 22.0$& $130.3\pm 20.4$& $143.8\pm 26.8$& $<0.001$ \\
DiaBP (mmHg)               & $82.9\pm 11.9$ & $82.2\pm 11.3$ & $87.2\pm 14.3$ & $<0.001$ \\
Total cholesterol (mg/dL)  & $237.0\pm 44.7$& $235.3\pm 43.8$& $246.3\pm 48.1$& $<0.001$ \\
BMI (kg/m$^2$)             & $25.8\pm 4.1$  & $25.7\pm 3.9$  & $26.6\pm 4.6$  & $<0.001$ \\
Glucose (mg/dL)            & $81.9\pm 23.8$ & $80.7\pm 19.0$ & $88.8\pm 40.8$ & $<0.001$ \\
Heart rate (bpm)           & $75.7\pm 12.0$ & $75.6\pm 11.9$ & $76.4\pm 12.1$ & $0.299$  \\
Male sex, $n$ (\%)         & 1,682 (44.5\%) & 1,362 (42.5\%) & 320 (55.7\%)   & $<0.001$ \\
Current smoker, $n$ (\%)   & 1,857 (49.2\%) & 1,561 (48.8\%) & 296 (51.6\%)   & $0.231$  \\
Diabetes mellitus, $n$ (\%)&   102  (2.7\%) &    66  (2.1\%) &  36  (6.3\%)   & $<0.001$ \\
Antihypertensive med., $n$ (\%)& 114 (3.0\%)& 77 (2.4\%)     &  37  (6.4\%)   & $<0.001$ \\
Prevalent hypertension, $n$ (\%)& 1,176 (31.1\%)& 885 (27.6\%)& 291 (50.7\%) & $<0.001$ \\
\bottomrule
\end{tabular}
\end{table}

\subsection*{Observational prediction model}

A multivariable logistic regression achieved an AUROC of 0.721
(5-fold cross-validation: $0.721 \pm 0.028$; average precision:
0.339; Brier score: 0.116).
The observational odds ratio per mmHg SysBP was
1.013 (95\% CI: 1.008--1.018; $p<0.001$).
Age (OR: 1.063), male sex (OR: 1.805), and current smoking
(OR: 1.477) were the strongest associated predictors.
These coefficients reflect the conditional distribution of CHD
given observed covariates and carry no causal interpretation.

\subsection*{Directed acyclic graph specification and evaluation}

\begin{figure}[H]
\centering
\includegraphics[width=0.9\linewidth]{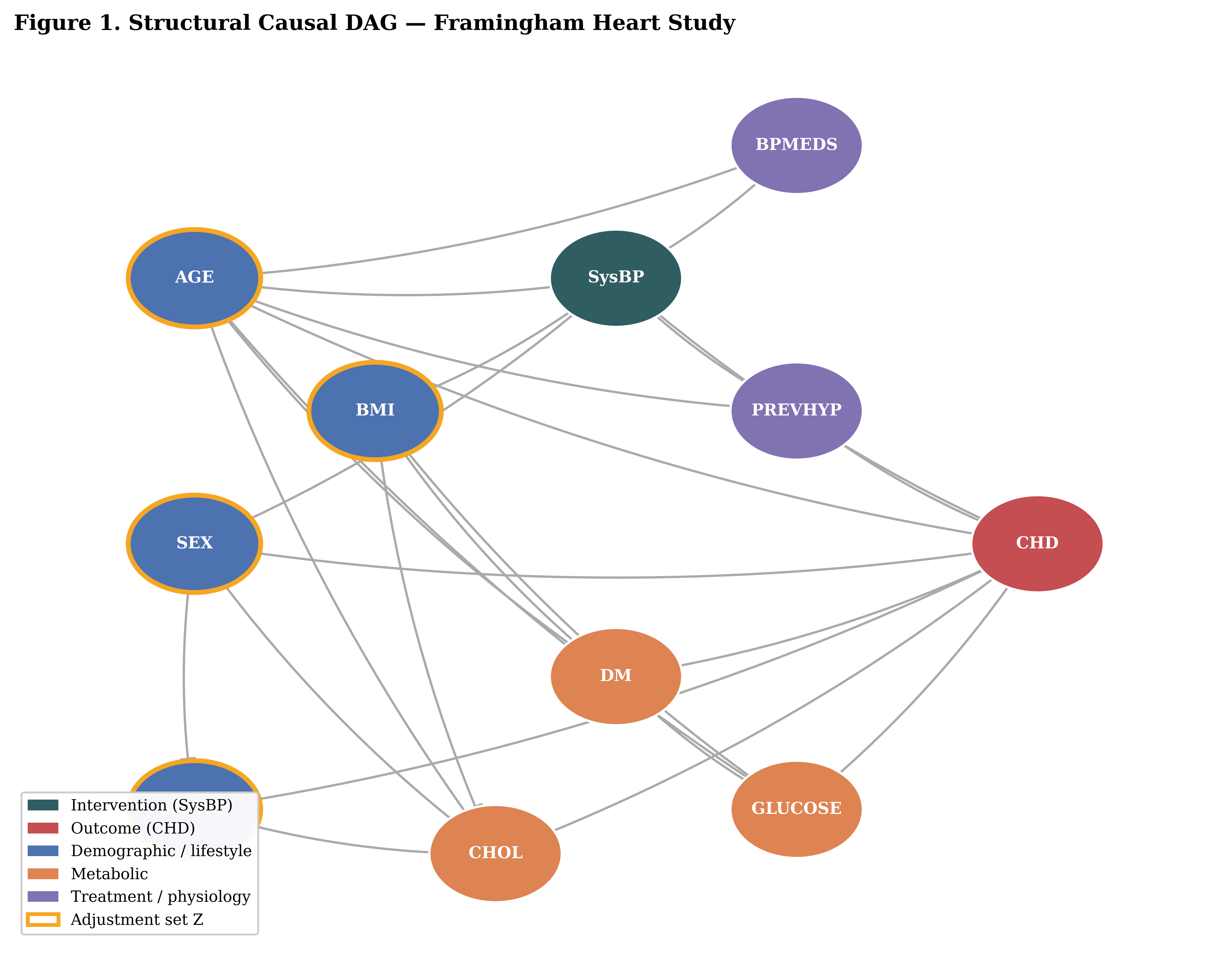}
\caption{\textbf{Corrected structural causal directed acyclic graph (DAG).}
Nodes represent Framingham Heart Study variables; directed edges
encode causal mechanisms supported by cardiovascular pathophysiology.
This DAG is a simplified representation and does not encode all biological
dependencies.
Four structural corrections differentiate this formulation from
prior approaches: removal of the cross-sectionally unjustified
smoking-to-SYSBP edge; addition of the biologically motivated
age-to-diabetes edge; representation of BPMEDS as a descendant of
SYSBP rather than a confounder; and exclusion of PREVHYP from the primary
adjustment set.
The back-door adjustment set
$Z=\{\mathrm{AGE},\mathrm{SEX\_MALE},\mathrm{BMI},\mathrm{CURSMOKE}\}$
is highlighted.}
\label{fig:dag}
\end{figure}

The corrected DAG (Figure~\ref{fig:dag}) encoded 11 nodes and 24
directed edges incorporating four biologically motivated structural
corrections.

\emph{Correction 1: Removal of CURSMOKE to SYSBP.}
Nicotine causes acute transient vasoconstriction~\cite{Halimi2002},
but in this cross-sectional cohort this does not constitute a
structural population-level causal path from smoking status to
sustained SysBP.
Retaining this edge creates an illegitimate back-door path through
an intermediate variable; the epidemiologically appropriate
structure routes smoking's CHD effect through cholesterol and via a
direct atherogenic mechanism.
\emph{Correction 2: Addition of AGE to DIABETES.}
Type~2 diabetes incidence increases sharply with
age~\cite{Gregg2014}.
Omitting this edge misrepresents age as an independent root cause
of CHD, failing to capture its role as a shared cause of both
SysBP elevation (through arterial stiffening) and
CHD (partially mediated through glucose dysregulation).

\emph{Correction 3: Reclassification of BPMEDS.}
Antihypertensive medication is prescribed because SysBP
is elevated; the causal arrow runs from SYSBP to BPMEDS.
Including BPMEDS in the back-door adjustment set introduces collider
bias, blocking part of the SYSBP to CHD effect pathway and
attenuating the true causal estimate~\cite{Pearl2009}.
BPMEDS was retained in the observational model as a precision
variable but was explicitly excluded from all causal adjustment sets.

\emph{Correction 4: Exclusion of PREVHYP from the primary
adjustment set.}
Prevalent hypertension is causally downstream of long-run SysBP.
Conditioning on it in the primary adjustment set would induce
over-adjustment bias; it was excluded from the primary causal model
and included only in a pre-specified sensitivity analysis.

We acknowledge that the DAG is a simplified representation of
biological reality.
Not all possible biological dependencies are encoded, and edges
were specified on the basis of cardiovascular pathophysiology rather
than formal elicitation.
The DAG should not be interpreted as a validated complete causal
map of the cardiovascular system.

Four conditional independence implications of the corrected DAG
were tested by partial correlation (Table~\ref{tab:S1}).
All four passed at $\alpha=0.05$:
$\mathrm{SEX}\perp\mathrm{GLUCOSE}\mid\{\mathrm{BMI},\mathrm{DIABETES}\}$
($r=-0.007$, $p=0.656$);
$\mathrm{BPMEDS}\perp\mathrm{TOTCHOL}\mid\{\mathrm{AGE},\mathrm{SYSBP}\}$
($r=+0.023$, $p=0.131$);
$\mathrm{BPMEDS}\perp\mathrm{GLUCOSE}\mid\{\mathrm{AGE},\mathrm{SYSBP}\}$
($r=+0.012$, $p=0.422$);
and $\mathrm{AGE}\perp\mathrm{BPMEDS}\mid\mathrm{SYSBP}$
($r=+0.025$, $p=0.109$).
The expected failure of
$\mathrm{SEX}\perp\mathrm{SYSBP}\mid\{\mathrm{AGE},\mathrm{BMI}\}$
($p<0.001$) correctly reflects the retained biologically established
direct effect of sex on SysBP~\cite{Vasan2001} and
corroborates rather than contradicts the specified graph.
Passage of all non-trivial testable implications is consistent with
the corrected DAG structure as a basis for causal identification,
although we cannot exclude misspecification of paths not directly
testable from these data.

\begin{table}[H]
\caption{\textbf{Supplementary Table S1.} Conditional independence
tests corresponding to testable implications of the corrected DAG.
Partial correlation coefficients computed by regressing out the
conditioning set via ordinary least squares.
$p>0.05$ is consistent with the null implication.
The failure of $\mathrm{SEX}\perp\mathrm{SYSBP}\mid\{\mathrm{AGE},\mathrm{BMI}\}$
is expected under the retained SEX to SYSBP edge and corroborates
rather than contradicts the DAG.}
\label{tab:S1}
\centering
\small
\begin{tabular}{lrrl}
\toprule
\textbf{Implication} & \textbf{Partial $r$} & \textbf{\textit{p}} & \textbf{Result} \\
\midrule
$\mathrm{SEX}\perp\mathrm{GLUCOSE}\mid\{\mathrm{BMI},\mathrm{DIABETES}\}$ & $-0.007$ & $0.656$ & PASS \\
$\mathrm{BPMEDS}\perp\mathrm{TOTCHOL}\mid\{\mathrm{AGE},\mathrm{SYSBP}\}$ & $+0.023$ & $0.131$ & PASS \\
$\mathrm{BPMEDS}\perp\mathrm{GLUCOSE}\mid\{\mathrm{AGE},\mathrm{SYSBP}\}$ & $+0.012$ & $0.422$ & PASS \\
$\mathrm{AGE}\perp\mathrm{BPMEDS}\mid\mathrm{SYSBP}$ & $+0.025$ & $0.109$ & PASS \\
$\mathrm{SEX}\perp\mathrm{SYSBP}\mid\{\mathrm{AGE},\mathrm{BMI}\}$ & --- & $<0.001$ & FAIL (expected; biologically valid direct edge retained) \\
\bottomrule
\end{tabular}
\end{table}

\subsection*{Average causal effect: the prediction--causation gap}

The formal causal estimand was
$E[Y\mid\mathrm{do}(\mathrm{SysBP}=s)]$,
identified by the back-door adjustment formula~\cite{Pearl2009}:
\begin{equation}
  E\!\left[Y \mid \mathrm{do}(\mathrm{SysBP}=s)\right]
  = \sum_{z} E\!\left[Y \mid \mathrm{SysBP}=s,\, Z=z\right] P(Z=z),
  \label{eq:backdoor}
\end{equation}
with adjustment set $Z=\{\mathrm{AGE},\mathrm{SEX\_MALE},\mathrm{BMI},\mathrm{CURSMOKE}\}$.
Total cholesterol and glucose were included in the outcome regression
as precision covariates; they are not back-door confounders of
SYSBP given $Z$, but reduce residual outcome variance without
introducing bias.
Three identifying assumptions were imposed: (i)~consistency
($Y_i = Y_i(\mathrm{SysBP}_i)$); (ii)~positivity
($P(\mathrm{SysBP}=s \mid Z=z)>0$ for all $z$ in the support of
$Z$ and all $s$ in the intervention range); and (iii)~conditional
ignorability ($Y(s)\perp\!\!\!\perp \mathrm{SysBP}\mid Z$ for all
$s$).

G-computation (standardisation) applied to the multiply-imputed
primary dataset ($n=4{,}240$) yielded an interventional 10-year
CHD risk of 14.11\% under
$\mathrm{do}(\mathrm{SysBP}=132~\mathrm{mmHg})$ and 10.71\%
under $\mathrm{do}(\mathrm{SysBP}=112~\mathrm{mmHg})$, giving:
\[
  \mathrm{ACE} = 3.40\%~\text{absolute risk reduction}~
  (95\%~\text{bootstrap CI: } 2.64\%\text{--}4.14\%;~
  \mathrm{RRR}=24.1\%).
\]
A complete-case sensitivity analysis ($n=3{,}776$) yielded
$\mathrm{ACE}=3.46\%$ (1.70\% difference from the primary
estimate), supporting robustness to the missing-data handling.

\begin{figure}[H]
\centering
\begin{subfigure}[t]{0.45\linewidth}
  \centering
  \includegraphics[width=\linewidth]{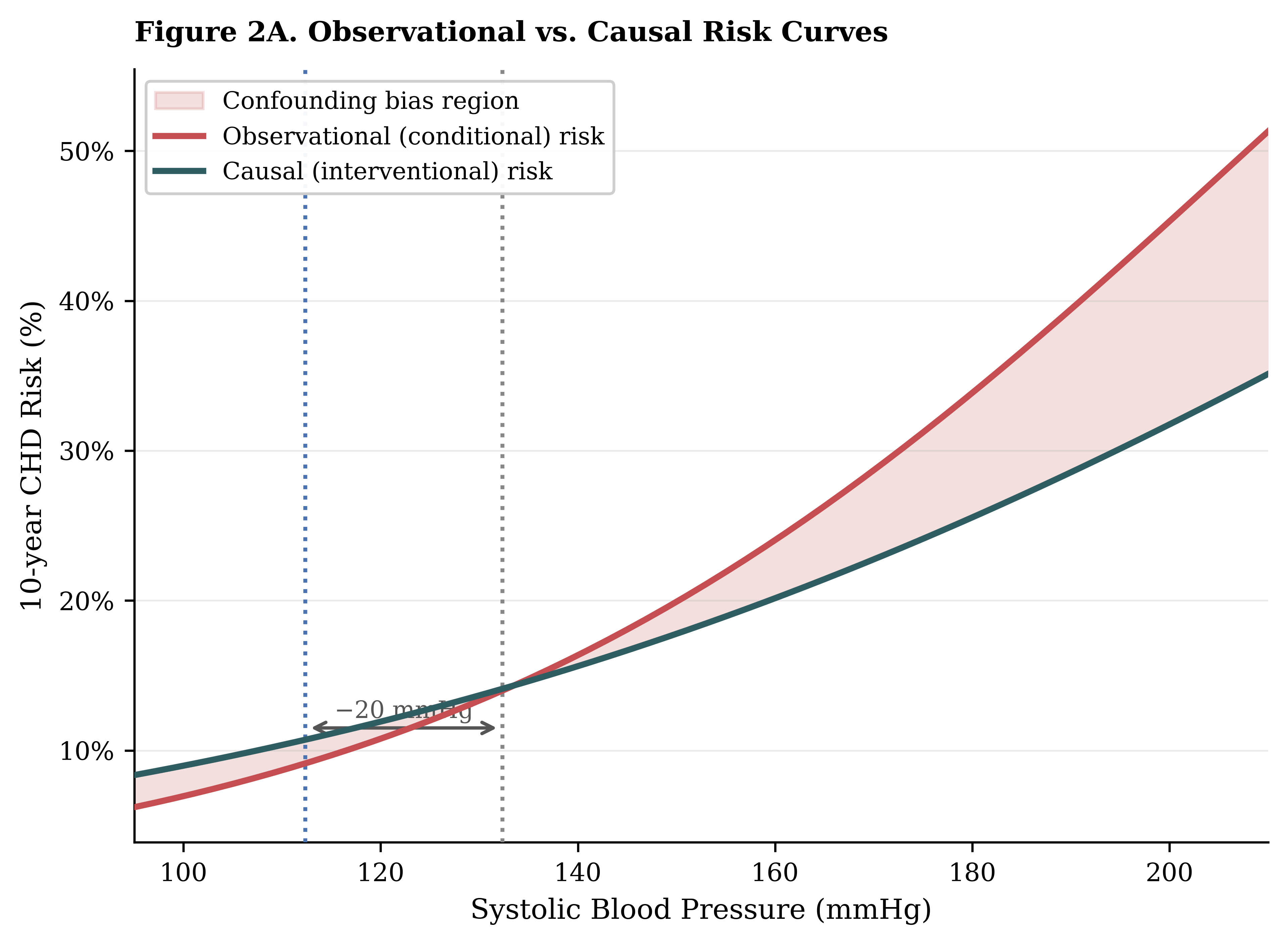}
  \caption{}
  \label{fig:bias_curve}
\end{subfigure}
\hfill
\begin{subfigure}[t]{0.45\linewidth}
  \centering
  \includegraphics[width=\linewidth]{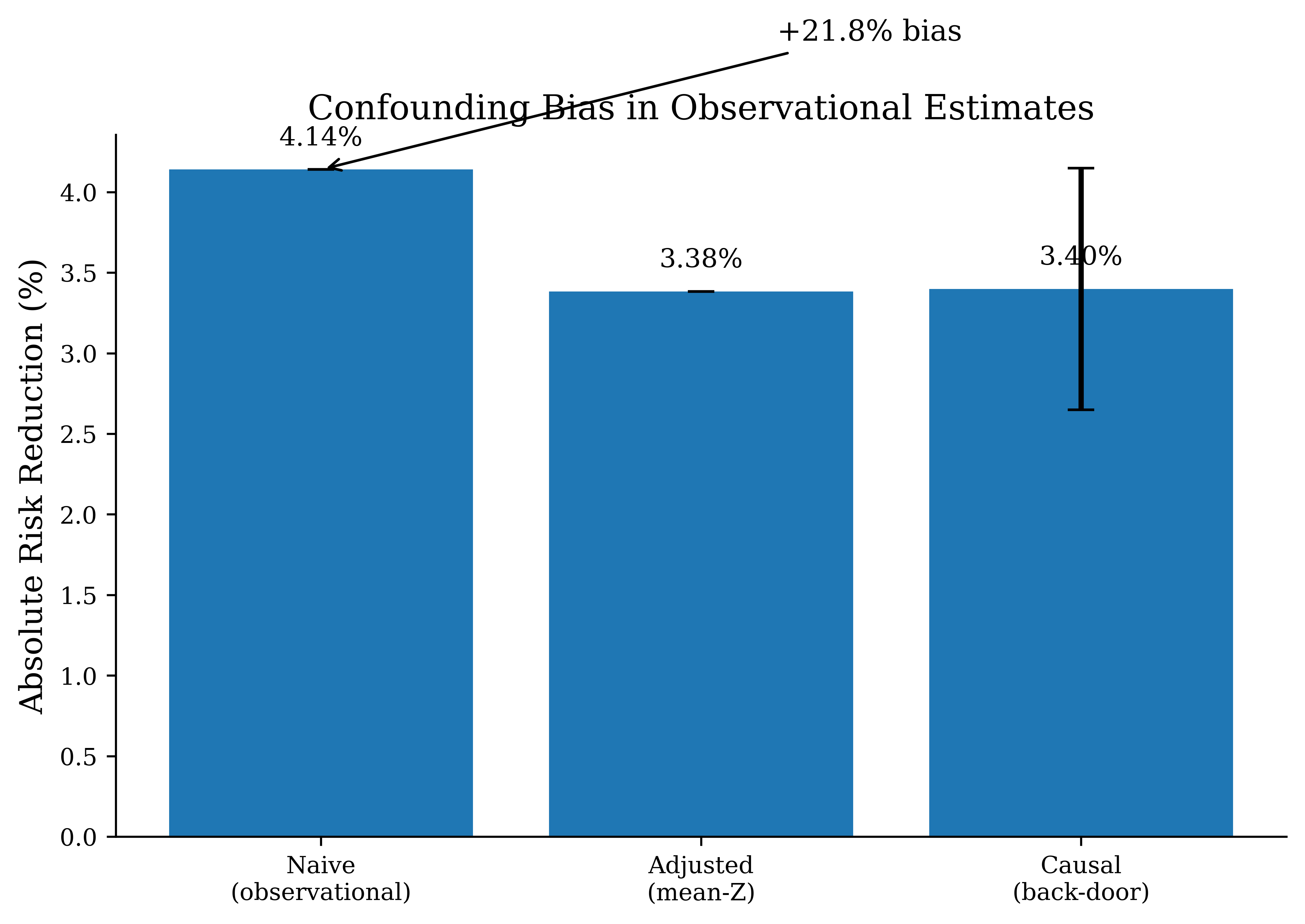}
  \caption{}
  \label{fig:bias_bar}
\end{subfigure}
\caption{\textbf{Bias in observational estimates of blood pressure benefit.}
\textbf{(A)} The observational (conditional) risk curve systematically exceeds
the causal (interventional) risk curve estimated by back-door
g-computation, reflecting confounding by age, BMI, and metabolic
comorbidity.
The shaded region represents the bias attributable to conflating
the conditional and interventional distributions, corresponding to
an approximate 21.8\% relative overestimation of absolute treatment
benefit at the primary intervention point.
\textbf{(B)} Bar comparison of absolute risk reduction estimates from the
naive observational approach, the mean-covariate plug-in adjusted
estimate, and the fully marginalised g-computation causal estimate.
Error bars represent 95\% bootstrap confidence intervals.}
\label{fig:bias}
\end{figure}

The naive observational estimate (comparing participants with SysBP
near 132 vs.\ 112~mmHg) was 4.14\%, standing 21.8\% above the causal
estimate in relative terms (log-odds inflation: $+35.7\%$),
driven primarily by the clustering of older age and higher BMI with
elevated SysBP.
The mean-$Z$ plug-in adjusted estimate was 3.38\%, confirming that
full marginalisation over the empirical covariate distribution
(rather than mere adjustment at covariate means) is necessary for
unbiased causal inference.
Figure~\ref{fig:bias} illustrates the magnitude of this divergence
between the observational and interventional estimates across the
range of SysBP values, and quantifies the bias
attributable to conflating the conditional and interventional
distributions.
These results are consistent with the hypothesis that observational
cardiovascular risk estimates overstate the expected absolute benefit
of antihypertensive treatment.

\subsection*{Triangulation: propensity score matching and
             inverse probability weighting}

A binary treatment contrast was defined by SysBP above versus below
the sample median (128~mmHg).
Sex-stratified nearest-neighbour PSM (caliper:
0.05; 2,169 matched pairs) achieved excellent post-matching
covariate balance (AGE SMD: $0.680\to0.053$; BMI SMD:
$0.598\to0.070$; CURSMOKE SMD: $0.211\to0.004$; SEX SMD:
$0.021\to0.000$).
The average treatment effect on the treated (ATT) from PSM
was 4.84\% (95\% bootstrap CI: 2.58\%--7.01\%);
the IPW population average treatment effect
(ATE) was 6.32\% (95\% CI: 3.90\%--8.75\%).

These estimates are larger than the back-door ACE, which is
expected: PSM and IPW
evaluate a binary contrast at the sample median, representing a more
extreme treatment contrast than the marginal 20~mmHg shift estimated
by g-computation.
Crucially, these three estimators rest on partially overlapping
assumptions and distinct identification strategies; their
directional consistency constitutes meaningful triangulation in
support of a causal interpretation of the primary estimate, and
does not imply that the estimates are directly comparable.

\begin{figure}[H]
\centering
\begin{subfigure}[t]{0.45\linewidth}
  \centering
  \includegraphics[width=\linewidth]{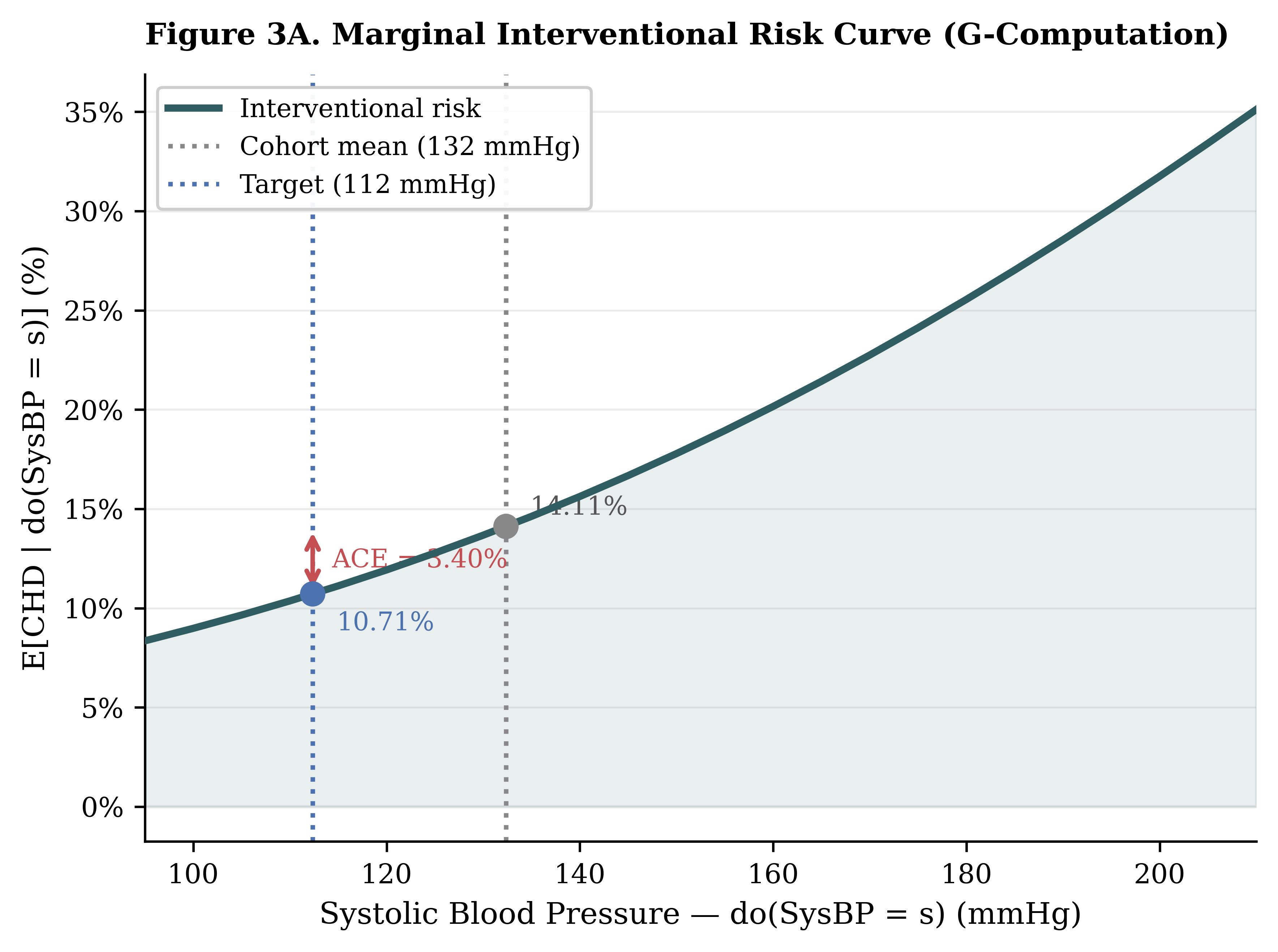}
  \caption{}
  \label{fig:gcomp_curve}
\end{subfigure}
\hfill
\begin{subfigure}[t]{0.45\linewidth}
  \centering
  \includegraphics[width=\linewidth]{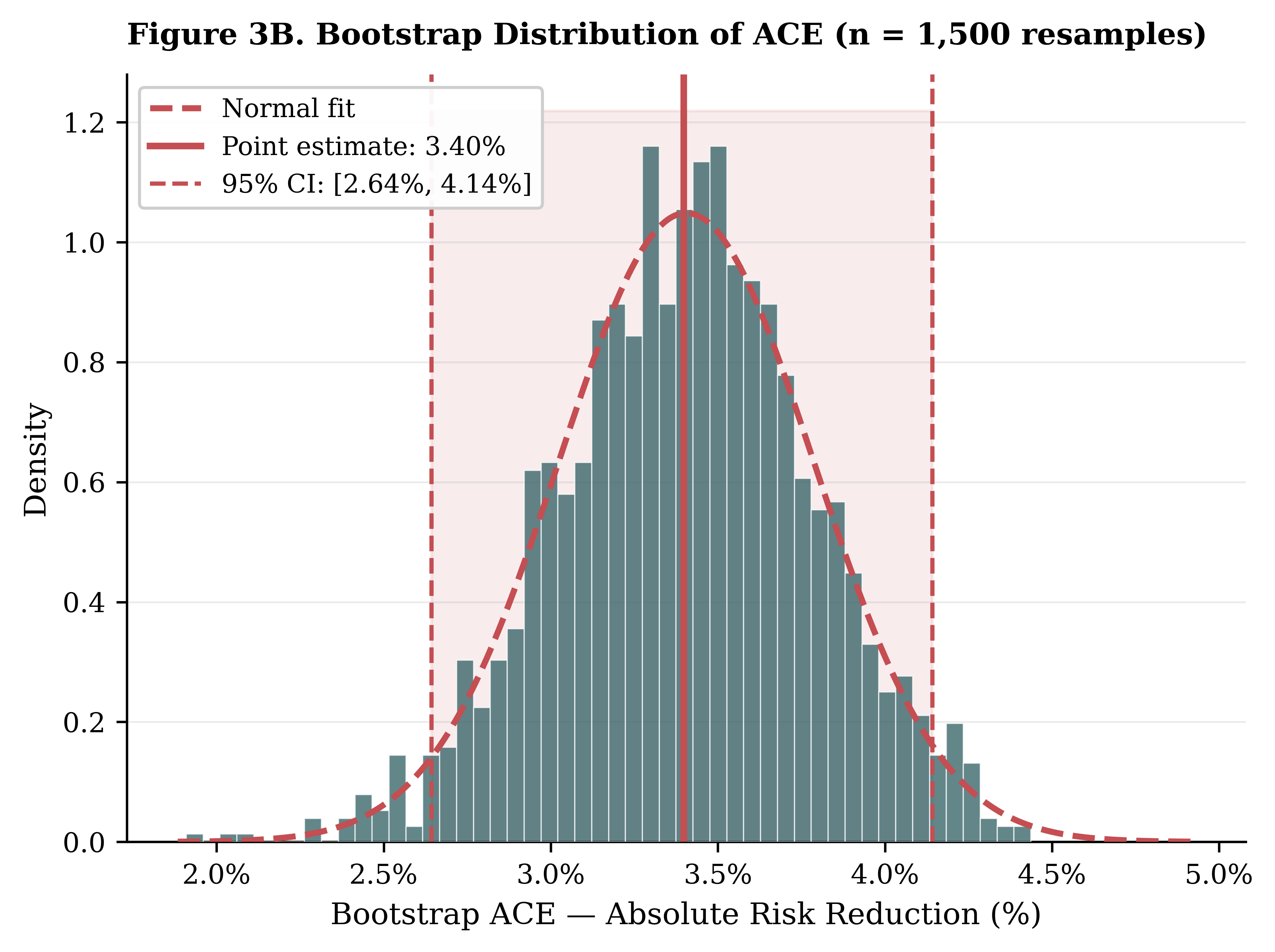}
  \caption{}
  \label{fig:bootstrap}
\end{subfigure}
\caption{\textbf{G-computation causal effect estimation.}
\textbf{(A)} Marginal interventional 10-year CHD risk as a function of
$\mathrm{do}(\mathrm{SysBP})$, estimated by back-door standardisation.
The grey dotted line marks the cohort mean (132.4~mmHg) and the green dotted
line marks the reduced target value (112.4~mmHg).
\textbf{(B)} Bootstrap distribution of the ACE for a 20~mmHg SysBP reduction
across 1,500 resamples.
The point estimate (3.40\%) and 95\% CI limits (2.64\%, 4.14\%) are marked
by vertical dashed lines.
Error bars represent the 2.5th and 97.5th percentiles of the bootstrap
distribution.}
\label{fig:ace}
\end{figure}

\subsection*{Refutation testing}

\begin{figure}[H]
\centering
\begin{subfigure}[t]{0.45\linewidth}
  \centering
  \includegraphics[width=\linewidth]{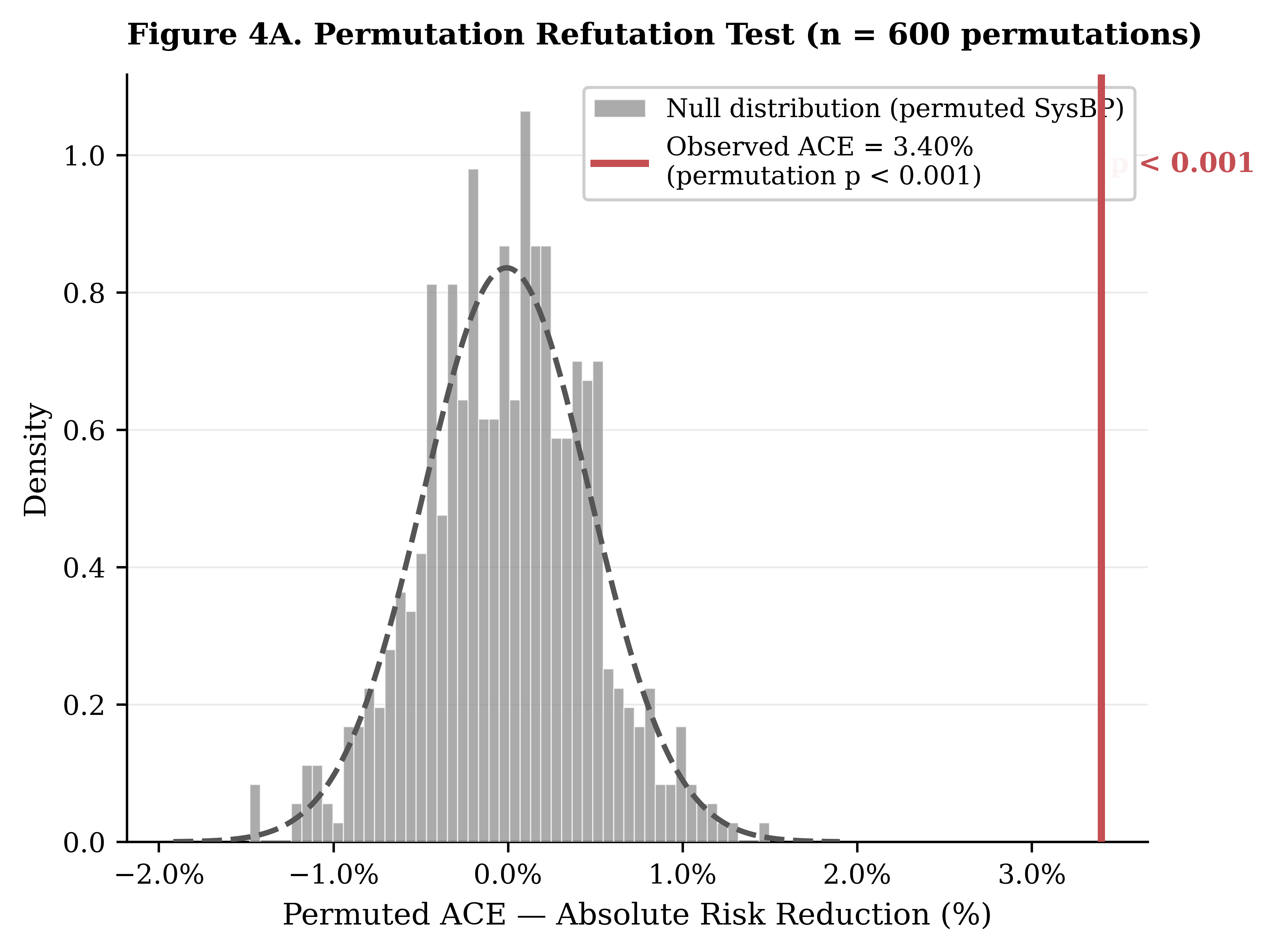}
  \caption{}
  \label{fig:permutation}
\end{subfigure}
\hfill
\begin{subfigure}[t]{0.45\linewidth}
  \centering
  \includegraphics[width=\linewidth]{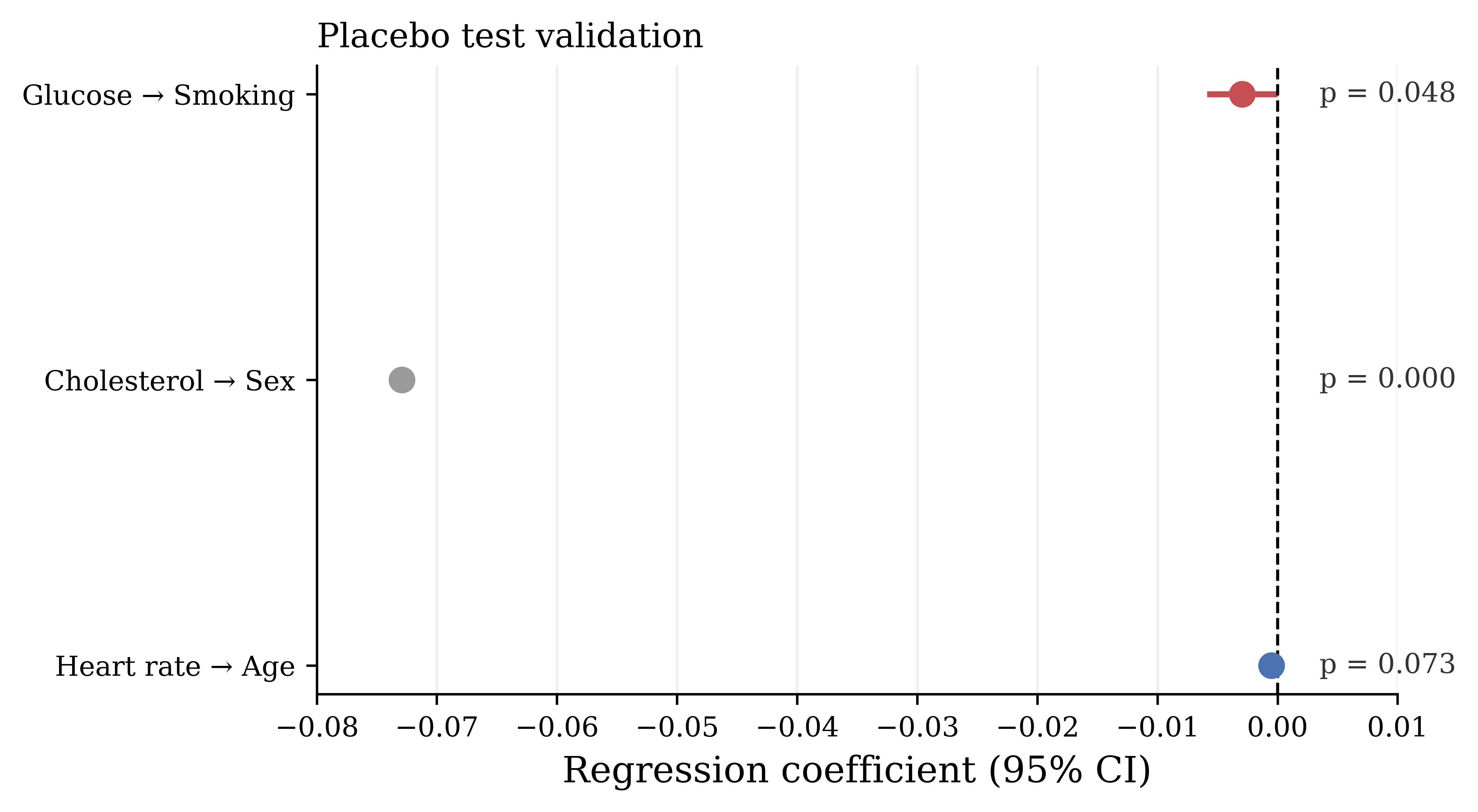}
  \caption{}
  \label{fig:placebo}
\end{subfigure}
\caption{\textbf{Validation and robustness of the causal estimate.}
\textbf{(A)} Null distribution of ACE estimates from 600 permutations of
the SysBP treatment vector.
The observed ACE (red vertical line, 3.40\%) lies far outside the
null distribution centred at approximately zero (permutation $p<0.001$),
rejecting the hypothesis that the effect arises from chance.
\textbf{(B)} Regression coefficients and 95\% confidence intervals for
placebo instruments.
Placebo~1 (heart rate to age) passed ($p=0.073$).
Placebo~3 (glucose to current smoking) failed ($p=0.048$), most likely
reflecting latent shared behavioural--metabolic causes not encoded
in the DAG; this indicates model simplification.
Placebo~2 (total cholesterol to sex) was pre-specified as invalid and
is shown for transparency only.
Error bars represent 95\% confidence intervals.}
\label{fig:refutation}
\end{figure}

The permutation refutation yielded a null distribution of ACEs
centred at $-0.000081$ ($\pm 0.0054$); the observed ACE fell in the
extreme tail ($p<0.001$ by permutation, Figure~\ref{fig:refutation}),
confirming that the estimated effect cannot be attributed to chance
or model artefact.

Three temporally or biologically impossible placebo instruments were
evaluated.
Placebo~1 (heart rate cannot causally precede age): coefficient $-0.000516$
(95\% CI: $[-0.001081,\ 0.000048]$, $p=0.073$), passed.
Placebo~3 (cross-sectional glucose cannot retroactively determine
historical smoking): coefficient $-0.002948$
(95\% CI: $[-0.005866,\ -0.000030]$, $p=0.048$), failed at the
boundary of $\alpha=0.05$.
Placebo~2 (total cholesterol to sex) failed due to a genuine
biological sex-to-cholesterol hormonal pathway encoded in the DAG; it
was pre-specified as invalid and is reported for transparency only.

We note explicitly that Placebo~3 failed.
The residual glucose--smoking association most likely reflects
latent shared causes, behavioural and metabolic clustering not
fully encoded in the DAG, rather than a direct causal path.
Similarly, the total cholesterol--sex pathway reflects established
hormonal--metabolic linkages.
These failures indicate model simplification rather than gross
misspecification, and do not invalidate the back-door identification
strategy for the primary estimand; however, they represent a
limitation that we acknowledge explicitly.

\subsection*{Conditional average treatment effects}

Heterogeneous treatment effects were estimated using R-Learner
(continuous treatment; $\tau$ per mmHg SysBP) and
T-Learner (binary treatment contrast at sample median).
Spearman correlation between R-Learner and T-Learner individual
estimates was 0.335, indicating directional consistency with
substantial noise, as expected for semiparametric metalearners on
observational data.

Statistically significant heterogeneity was detected across age
strata (Kruskal--Wallis $p<0.001$) and diabetes status
(Mann--Whitney $p<0.001$), and across sex (Mann--Whitney $p<0.001$),
as shown in Table~\ref{tab:cate}.
Age was the most clearly stratifying variable: participants younger
than 45 years showed an estimated implied absolute risk reduction
(ARR) of 2.22\% (95\% bootstrap CI: 1.87\%--2.59\%), while those
aged 45--54, 55--64, and $\geq 65$ showed ARRs of approximately
4.49\%, 4.76\%, and 3.74\%, respectively.
This age gradient is consistent with guideline recommendations
derived from committee consensus~\cite{Whelton2018} but is
here derived from direct causal estimation.

\textbf{Diabetic subgroup estimates must be interpreted with extreme
caution.}
The diabetic subgroup comprised only $n=109$ participants.
A minimum detectable effect of 0.00240 per mmHg (at 80\% power)
substantially exceeds the observed $\tau$ of 0.00087, indicating
that this subgroup is underpowered for reliable effect detection.
The 95\% confidence interval for the diabetic subgroup ARR
($-4.94\%$ to $+2.08\%$) crosses zero and is wide.
We cannot make reliable inference about differential benefit in
diabetic patients from this cohort; no claim that diabetic
participants derive greater benefit from blood pressure reduction
should be drawn from these data without replication in a larger or
enriched diabetic sample.
All heterogeneity results should be interpreted cautiously.

\begin{table}[H]
\caption{\textbf{Conditional average treatment effects (CATE) by
pre-specified subgroup.}
Estimates from the R-Learner with gradient-boosted nuisance models;
treatment is continuous SysBP.
Implied ARR is expressed for a $-20$~mmHg
intervention ($\mathrm{ARR} = 20\times\bar{\tau}$).
95\% confidence intervals from 500-iteration bootstrap resampling
of subgroup means.
\textit{Diabetic subgroup estimates ($n=109$) are underpowered
(minimum detectable effect $=0.00240$ per mmHg vs.\ observed
$\tau=0.00087$) and should not be used for inference.}
All heterogeneity results should be interpreted with caution.}
\label{tab:cate}
\centering
\small
\begin{tabular}{lrcccc}
\toprule
\textbf{Subgroup} & $n$ & \textbf{$\bar{\tau}$ (per mmHg)} &
\textbf{Implied ARR} & \textbf{95\% CI} & \textbf{\textit{p}} \\
\midrule
\multicolumn{6}{l}{\textit{Sex (Mann--Whitney $U$)}} \\
Male   & 1,820 & $0.00321$ & $6.42\%$ & $[5.97\%,\ 6.87\%]$ & \multirow{2}{*}{$<0.001$} \\
Female & 2,420 & $0.00082$ & $1.64\%$ & $[1.25\%,\ 2.02\%]$ & \\
\midrule
\multicolumn{6}{l}{\textit{Diabetes status (Mann--Whitney $U$)}} \\
Diabetic$^{\dagger}$    & 109   & $-0.00087$ & $-1.73\%$ & $[-4.94\%,\ +2.08\%]$ & \multirow{2}{*}{$<0.001$} \\
Non-diabetic & 4,131 & $0.00192$ & $3.83\%$ & $[3.55\%,\ 4.11\%]$ & \\
\midrule
\multicolumn{6}{l}{\textit{Age strata (Kruskal--Wallis)}} \\
$<$45 years  & 1,589 & $0.00111$ & $2.22\%$ & $[1.87\%,\ 2.59\%]$ & \multirow{4}{*}{$<0.001$} \\
45--54 years & 1,479 & $0.00225$ & $4.49\%$ & $[3.98\%,\ 4.97\%]$ & \\
55--64 years & 1,062 & $0.00238$ & $4.76\%$ & $[4.11\%,\ 5.41\%]$ & \\
$\geq$65 years & 110 & $0.00187$ & $3.74\%$ & $[0.65\%,\ 6.45\%]$ & \\
\bottomrule
\end{tabular}
\vspace{2pt}
\footnotesize{$^{\dagger}$Underpowered; no reliable inference possible without replication.}
\end{table}

\subsection*{Sensitivity analysis: E-values}

\begin{figure}[H]
\centering
\begin{subfigure}[t]{0.45\linewidth}
  \centering
  \includegraphics[width=\linewidth]{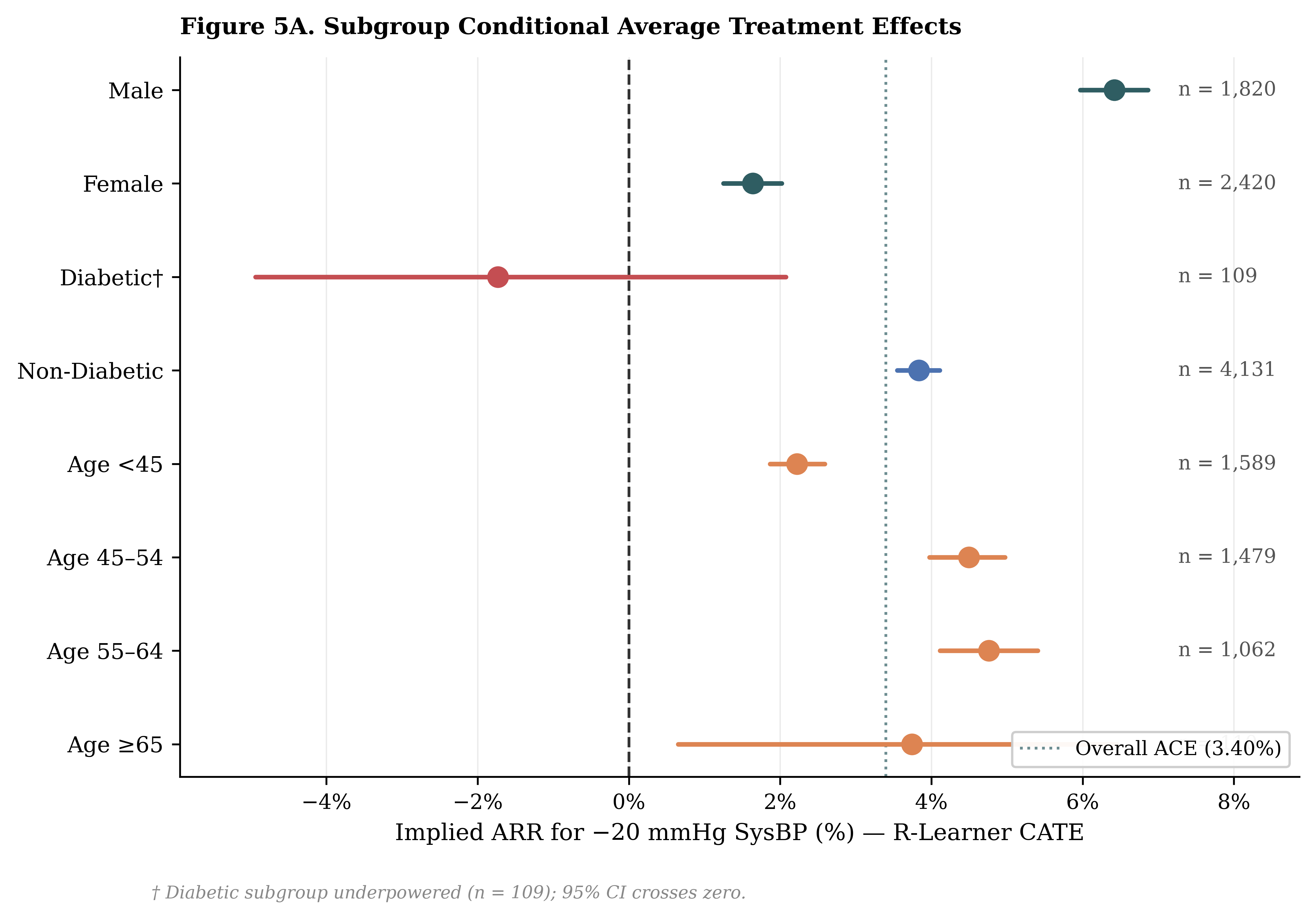}
  \caption{}
  \label{fig:cate}
\end{subfigure}
\hfill
\begin{subfigure}[t]{0.45\linewidth}
  \centering
  \includegraphics[width=\linewidth]{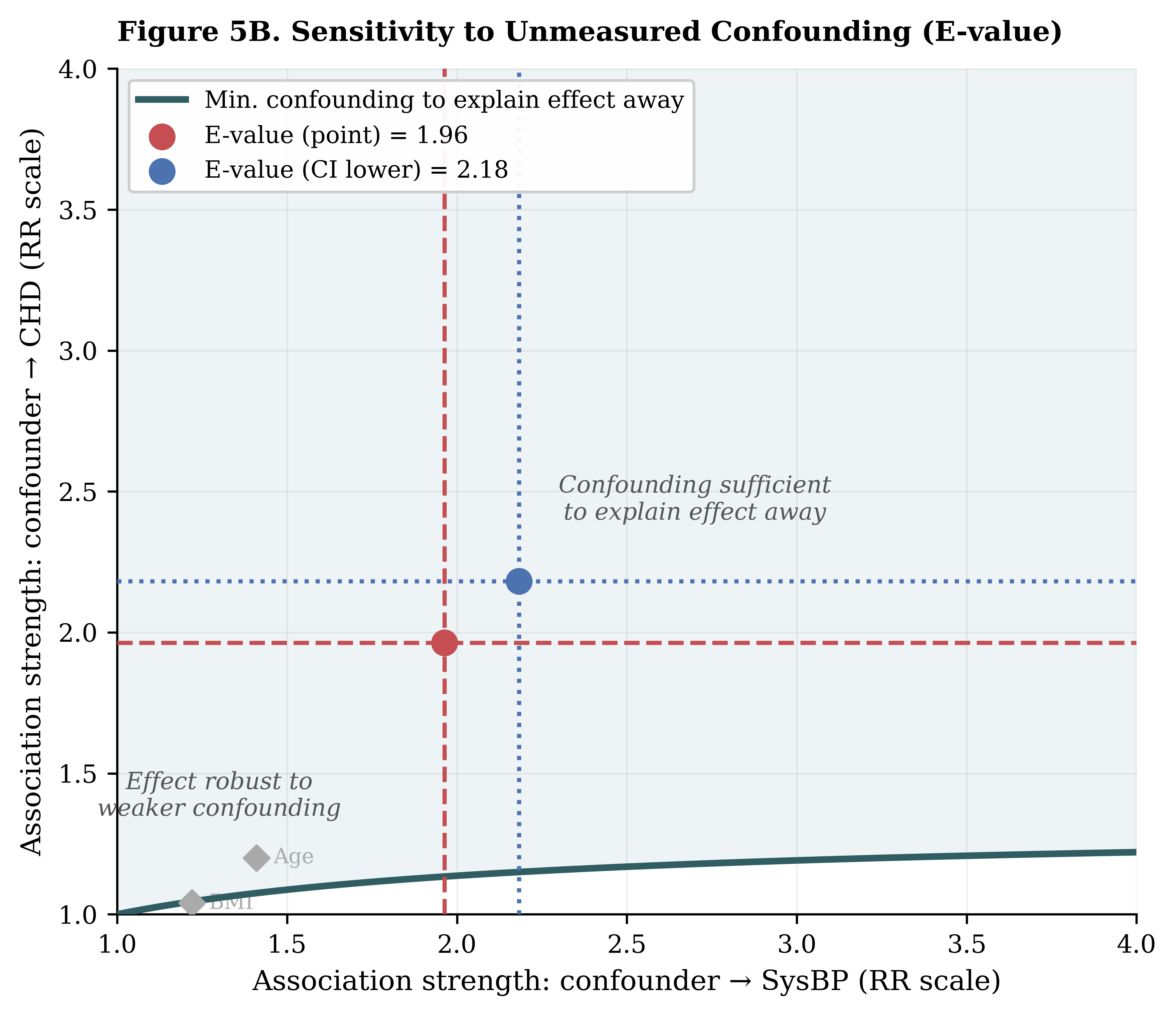}
  \caption{}
  \label{fig:evalue}
\end{subfigure}
\caption{\textbf{Heterogeneity in treatment effects and sensitivity to unmeasured confounding.}
\textbf{(A)} Implied ARR for a 20~mmHg SysBP reduction
from the continuous-treatment R-Learner, expressed per subgroup.
Error bars represent bootstrap standard errors.
The diabetic subgroup ($n=109$, marked with $\dagger$) is underpowered
(minimum detectable effect greater than observed $\tau$); its estimate
should not be interpreted as reliable.
All heterogeneity findings are exploratory and require replication.
\textbf{(B)} Sensitivity curve for unmeasured confounding.
The curve represents the minimum strength of association that an
unmeasured confounder would need with both SysBP
and CHD to explain away the observed causal effect.
The E-value (1.96) and CI lower-bound E-value (2.18) indicate that
only moderately strong confounding could negate the estimated effect,
and that this threshold exceeds the confounding magnitude observed
for any measured covariate in the dataset.}
\label{fig:heterogeneity}
\end{figure}

The interventional risk ratio was 1.317.
The E-value for the point estimate was 1.96; for the confidence
interval lower bound it was 2.18~\cite{VanderWeele2017},
exceeding the confounding strength of every measured covariate
(age: $R^2_{\mathrm{SysBP}}=0.108$, $R^2_{\mathrm{CHD}}=0.043$;
BMI: $R^2_{\mathrm{SysBP}}=0.083$, $R^2_{\mathrm{CHD}}=0.002$).
This suggests the estimated causal effect is robust to unmeasured
confounding of a magnitude not observed among the measured risk
factors, though we acknowledge this does not rule out unmeasured
confounders of sufficient strength.
E-values declined to 1.88 and 1.62 for 15 and 10~mmHg reductions
respectively; the 20~mmHg intervention magnitude is the only one
achieving CI-lower-bound robustness above 2.0.
Figure~\ref{fig:heterogeneity} displays the CATE subgroup effects
and the sensitivity curve showing the minimum confounding strength
required to nullify the estimated effect across the range of
plausible unmeasured confounder associations.

\begin{table}[H]
\caption{\textbf{Supplementary Table S2.} E-values by intervention
magnitude~\cite{VanderWeele2017}.
E-values represent the minimum risk-ratio-scale association that an
unmeasured confounder would need to have with both SYSBP and CHD to
fully explain away the estimated causal effect.
CI lower-bound E-values $>2.0$ indicate high robustness; this
threshold is met only at the 20~mmHg magnitude.}
\label{tab:S2}
\centering
\small
\begin{tabular}{lcccc}
\toprule
\textbf{Intervention} & \textbf{ACE (\%)} & \textbf{Causal RR} &
\textbf{E-value (point)} & \textbf{E-value (CI lower)} \\
\midrule
$-20$~mmHg SysBP & 3.40 & 1.317 & 1.96 & 2.18 $\checkmark$ \\
$-15$~mmHg SysBP & 2.55 & 1.233 & 1.74 & 1.88 \\
$-10$~mmHg SysBP & 1.70 & 1.150 & 1.53 & 1.62 \\
\bottomrule
\end{tabular}
\end{table}

% =======================================================
\section*{Discussion}
% =======================================================

This study suggests that standard observational cardiovascular risk
tools may overestimate the absolute benefit of SysBP
reduction by approximately one-fifth.
G-computation applied to the Framingham Heart Study Offspring Cohort
yields an ACE of 3.40\% (95\% CI: 2.64\%--4.14\%) absolute risk
reduction for a 20~mmHg decrease in SysBP, compared
with a naive observational estimate of 4.14\%.
The relative discrepancy of 21.8\% arises from the systematic
clustering of risk factors, older age, higher BMI, and metabolic
dysregulation, with elevated blood pressure.
Antihypertensive therapy modifies blood pressure, not the entire
comorbidity profile; the causal estimate captures only the
blood-pressure-specific component of absolute risk.

The clinical implications of this finding, if confirmed, are
consequential.
A clinician relying on an observational risk score to estimate the
benefit of antihypertensive treatment would, under this framework,
systematically overestimate the expected absolute risk reduction by
approximately one-fifth.
For patients at moderate absolute risk, this overestimation
could influence marginal prescribing decisions where the true causal
benefit may be below a clinically meaningful threshold.
Conversely, the treatment-effect heterogeneity analysis suggests,
with the caveats noted above, that older patients may derive greater
absolute benefit than younger individuals, which is consistent with
the higher baseline event rates in this age group.
These observations support the value of estimating interventional
rather than associational effects in cardiovascular risk prediction,
and suggest that causal methods could contribute to improved risk
stratification.
We stress that these are observational, cross-sectional estimates
and should be interpreted as hypothesis-generating rather than
definitive.

To illustrate the clinical consequence concretely, consider a 
55-year-old male non-smoker with a BMI of 27~kg/m$^2$ and a 
SysBP of 148~mmHg whose ten-year CHD risk under the observational 
Framingham Risk Score is estimated at 12\%. 
At a commonly applied 10\% prescribing threshold, the observational 
estimate would suggest antihypertensive treatment is clearly 
warranted.
Applying the 21.8\% relative correction derived here, the 
corresponding causal estimate of absolute benefit from a 
20~mmHg reduction falls to approximately 9.8\%---below the 
treatment threshold.
This is not an argument against antihypertensive treatment in 
such patients, for whom clinical judgment and total risk burden 
remain paramount; it is an argument that the quantity informing 
the prescribing decision should be an interventional estimate, 
not a conditional association.
For patients clustered near any prescribing threshold, the 
systematic one-fifth overestimation documented here represents 
a structurally embedded bias that causal methods are positioned 
to correct.

The estimated ACE of 3.40\% over ten years is directionally
consistent with the SPRINT trial~\cite{SPRINT2015}, which reported
a 1.65\% absolute reduction in major cardiovascular events at three
years targeting SysBP $<120$ versus $<140$~mmHg in non-diabetic
high-risk adults, and with the meta-analysis by Ettehad and
colleagues~\cite{Ettehad2016} reporting greater absolute benefit at
higher baseline risk.
Differences in magnitude are plausible given the longer time
horizon, the broader population, and the use of the cohort mean
SysBP as the baseline intervention point.

The methodological contribution of this work is the integration
of DAG evaluation, g-computation, sex-stratified PSM,
IPW, rigorous refutation testing,
and semiparametric metalearner CATE estimation within a single
reproducible pipeline applied to a publicly available dataset.
The four DAG corrections introduced here transfer directly to any
causal analysis of cardiovascular data in which antihypertensive
medication and prevalent hypertension are measured.
In particular, conditioning on antihypertensive medication as a
confounder, rather than correctly treating it as a descendant of
blood pressure, introduces collider bias that attenuates causal
estimates; this error appears in several published causal analyses
of cardiovascular data.

\textbf{Limitations.}
We acknowledge several important limitations.

First, the identifying assumption of no unmeasured confounding
conditional on $Z$ is unverifiable from observational data.
Unmeasured determinants of both SysBP and CHD
risk, including dietary patterns, physical activity, socioeconomic position,
and genetic predisposition, could bias the ACE in either direction,
notwithstanding the E-value analysis.
We cannot exclude residual confounding of meaningful magnitude.

Second, the DAG is a simplified representation of the cardiovascular
system.
The failures of Placebo~3 (glucose to current smoking) and the
total cholesterol--sex instrument indicate that latent shared causes not
encoded in the DAG are present.
These failures do not invalidate the primary identification
strategy, but they indicate model simplification and suggest that
the true causal structure may be more complex than the specified DAG.

Third, the analysis is cross-sectional at baseline.
The ten-year follow-up period introduces time-varying confounding
not addressed by the static DAG structure.
A marginal structural model with time-varying IPW
would be required for a causal analysis of sustained blood
pressure trajectories.
The static DAG likely understates cumulative exposure in high-risk
individuals, suggesting that the g-computation ACE may be
conservative.

Fourth, the SYSBP--BPMEDS feedback loop is approximated by a static
DAG; an instrumental variable design using a genetic instrument for
blood pressure could resolve this structural simplification.

Fifth, diabetic subgroup CATE estimates ($n=109$) are underpowered
and unreliable.
The point estimate and confidence interval for this subgroup should
not be interpreted as evidence of differential benefit.

Sixth, the Framingham Heart Study Offspring Cohort is a largely
white, community-based New England sample enrolled in the mid-20th
century.
Generalisation of these estimates to contemporary, ethnically
diverse, or higher-risk clinical populations should be treated with
caution.

Seventh, we cannot exclude measurement error in self-reported
variables (smoking status) or single-occasion blood pressure
measurements, which may introduce non-differential misclassification
that could bias the ACE toward the null.

Eighth, missing-data handling relied on iterative imputation under
a missing-at-random assumption; the MCAR assumption was supported
for most variables but violated for BMI ($\chi^2$, $p<0.001$).

In summary, this study is consistent with the hypothesis that
standard observational cardiovascular risk tools overestimate the
causal benefit of blood pressure reduction by approximately
one-fifth.
Future work should prioritise validation against individual-
participant data from randomised trials (SPRINT, ACCORD-BP), extension
to the longitudinal Framingham data using marginal structural models,
and replication in larger and more diverse cohorts.

% =======================================================
\section*{Methods}
% =======================================================

\subsection*{Dataset and study population}

The Framingham Heart Study (FHS) Offspring Cohort is a longitudinal
cardiovascular epidemiology study established in 1948~\cite{Kannel1961}.
The analytical dataset comprised 4,240 participants with 21 baseline
variables including systolic and diastolic blood pressure, total
serum cholesterol, fasting glucose, body mass index (BMI), age, sex,
current smoking status, antihypertensive medication use (BPMEDS),
prevalent hypertension (PREVHYP), diabetes mellitus, and heart rate.
The primary outcome was 10-year incident CHD, a composite of
myocardial infarction, angina pectoris, coronary insufficiency, and
coronary death, with a raw event rate of 15.2\% (644 events).
The data are publicly available through BioLINCC
(\url{https://biolincc.nhlbi.nih.gov}) and were used under the
terms of the data use agreement.

\subsection*{Data preprocessing and missing data}

Missing data rates were: glucose (9.15\%), education (2.48\%),
BPMEDS (1.25\%), total cholesterol (1.18\%), cigarettes per day
(0.68\%), and BMI (0.45\%).
A chi-squared test of the missing-completely-at-random (MCAR)
hypothesis, conditional on CHD outcome, was conducted for each
variable.
All variables except BMI supported MCAR ($p>0.05$); BMI showed
evidence against MCAR ($p<0.001$)~\cite{Little2019}.
Primary analyses used iterative multiple imputation
(scikit-learn \texttt{IterativeImputer}; 10 iterations) on the full
cohort ($n=4{,}240$).
Sensitivity analyses on the complete-case sample ($n=3{,}776$;
89.1\% of the cohort) produced estimates within 1.70\% of the
primary analysis, supporting robustness.

\subsection*{Observational prediction model}

A multivariable logistic regression was fitted with predictors:
SysBP, total cholesterol, glucose, age, sex, current smoking,
diabetes, prevalent hypertension, BMI, and antihypertensive
medication use.
Model performance was assessed by AUROC, average precision, and
Brier score under 5-fold stratified cross-validation.
Odds ratios from this model represent associations only and were
not interpreted as causal effects.

\subsection*{Causal DAG specification and identification}

We specified a structural causal model following Pearl's
framework~\cite{Pearl2009}, encoding biological mechanisms as
non-parametric structural equations.
The DAG encodes 11 nodes and 24 directed edges.
Four biologically motivated corrections to prior formulations are
described in the Results.

The formal causal estimand was
$E[Y\mid\mathrm{do}(\mathrm{SysBP}=s)]$,
identified by the back-door adjustment formula (\ref{eq:backdoor}),
with adjustment set
$Z=\{\mathrm{AGE},\mathrm{SEX\_MALE},\mathrm{BMI},\mathrm{CURSMOKE}\}$.
We verified that $Z$ satisfies the back-door criterion: it blocks
all back-door paths from SYSBP to CHD, contains no descendants of
SYSBP, and renders potential outcomes conditionally independent of
observed treatment given $Z$.
Total cholesterol and glucose were included in the outcome regression
as precision covariates.

Three identifying assumptions were imposed:
(i)~\emph{Consistency:} $Y_i = Y_i(\mathrm{SysBP}_i)$;
(ii)~\emph{Positivity:} $P(\mathrm{SysBP}=s\mid Z=z)>0$ for all
$z$ in the support of $Z$ and all $s$ in the intervention range;
(iii)~\emph{Conditional ignorability (exchangeability):}
$Y(s)\perp\!\!\!\perp \mathrm{SysBP}\mid Z$ for all $s$, that is,
no unmeasured confounding given $Z$.

DAG testable implications were evaluated by partial correlation,
regressing out conditioning sets via ordinary least squares.

\subsection*{G-computation and average causal effect estimation}

The interventional risk
$E[Y\mid\mathrm{do}(\mathrm{SysBP}=s)]$ was estimated by
g-computation (standardisation)~\cite{Hernan2020}.
A logistic outcome model was fitted:
\begin{align}
  \mathrm{logit}\,P(\mathrm{CHD}_{10}=1) = &\;
  \beta_0 + \beta_1\,\mathrm{SYSBP} + \boldsymbol{\beta}_Z^\top Z
  \notag\\
  &+ \gamma_1\,\mathrm{TOTCHOL} + \gamma_2\,\mathrm{GLUCOSE}
  + \gamma_3\,\mathrm{DIABETES}.
  \label{eq:outcome}
\end{align}
For each intervention value $s$, the full dataset was replicated
with SysBP set to $s$ for every individual, and the marginal
interventional risk was computed as the mean predicted probability
over the empirical distribution of $Z$.
The ACE was the difference in interventional risks at the cohort
mean SysBP (132.4~mmHg) versus 20~mmHg below (112.4~mmHg).
Bootstrap confidence intervals (95\%) used 1,500 resamples with the
percentile method, applied to the complete estimation procedure
including model fitting.

\subsection*{Propensity score matching and inverse probability
             weighting}

For PSM, a binary treatment variable was
defined by SysBP above versus below the sample median (128~mmHg).
Propensity scores were estimated by logistic regression on $Z$.
Sex-stratified nearest-neighbour matching was performed (exact
matching on sex; caliper 0.05 on propensity score scale), yielding
2,169 matched pairs.
Post-matching balance was verified by standardised mean differences.
The ATT was estimated from matched outcome differences with
800-iteration bootstrap confidence intervals.

For IPW, stabilised Horvitz--Thompson
weights were constructed and trimmed at the 98th percentile.
The ATE was estimated as the weighted mean outcome difference with
800-iteration bootstrap confidence intervals.

It is important to note that g-computation, PSM, and IPW estimate different
estimands (population ATE vs.\ ATT vs.\ ATT at a binary contrast
vs.\ marginal ATE at a continuous shift) and are \emph{not} directly
comparable.
Their directional consistency nonetheless provides triangulation
evidence for a causal interpretation.

\subsection*{Refutation testing}

Two refutation strategies were employed.
First, three temporally or biologically impossible placebo
instruments were evaluated: (i)~heart rate cannot
causally precede age; (ii)~serum cholesterol cannot determine biological sex,
retained for transparency despite pre-specified invalidity due to the
sex-to-cholesterol hormonal path; (iii)~cross-sectional glucose cannot
retroactively determine historical smoking behaviour.
Second, the SYSBP vector was permuted across 600 iterations to
construct a null distribution for the ACE.

\subsection*{Conditional average treatment effect estimation}

CATE was estimated using R-Learner~\cite{Nie2021} and
T-Learner~\cite{Kunzel2019} metalearners.
The R-Learner estimated nuisance functions for the conditional
outcome and the conditional mean of SYSBP given covariates by
5-fold cross-fitting with gradient boosting machines
(200 estimators; maximum depth 3; learning rate 0.05).
The pseudo-outcome was computed for observations with
$|T_i - \hat{e}(X_i)|>2.0$, clipped to the [5th, 95th] percentile,
and a gradient-boosted CATE model was fitted by regressing
pseudo-outcomes on covariates weighted by the squared treatment
residual.
The T-Learner fitted separate outcome models to treated and control
subgroups and estimated CATE as their predicted probability
difference.
Subgroup CATE means were accompanied by 500-iteration bootstrap
95\% confidence intervals.
Heterogeneity across subgroups was tested by Mann--Whitney $U$
(sex, diabetes) and Kruskal--Wallis (age strata) tests.

\subsection*{Sensitivity analysis}

Sensitivity to unmeasured confounding was quantified by
E-values~\cite{VanderWeele2017}:
\[
  E\text{-value} = \mathrm{RR} + \sqrt{\mathrm{RR}(\mathrm{RR}-1)}.
\]
Partial $R^2$ values for each measured confounder with respect to
SYSBP and CHD provided empirical benchmarks for the required
confounding magnitude.

All analyses were implemented in Python 3.11 using NumPy~1.26.4,
Pandas~3.0.1, statsmodels, and scikit-learn.
Random seeds were fixed at 42 throughout.

% =======================================================
\section*{Data availability}
% =======================================================

All analyses were performed on the Framingham Heart Study
public-use dataset, freely available through the National Heart,
Lung, and Blood Institute's BioLINCC repository
(\url{https://biolincc.nhlbi.nih.gov}).
No new data were generated or collected for this study.

% =======================================================
\section*{Code availability}
% =======================================================

Full analysis code (Python 3.11) is available at
\url{https://github.com/Suchibrata-Patra/bp-causal-inference-framingham/}.
The code includes all preprocessing, DAG specification and testing,
causal effect estimation, refutation testing, and figure generation
steps, and is sufficient for complete reproduction of all results reported.

% =======================================================
\section*{Ethics statement}
% =======================================================

This study used de-identified, publicly available data from the
Framingham Heart Study obtained through an approved data use
agreement with BioLINCC/NHLBI.
No new human participant data were collected.
No additional ethical approval was required under the institutional
review framework applicable to the analysis of publicly available
de-identified datasets.

% =======================================================
\section*{Funding}
% =======================================================
This research did not receive external funding.

% =======================================================
\section*{Author contributions}

Suchibrata Patra conceived the study, developed the causal framework, performed all data analysis, implemented the computational pipeline, and wrote the manuscript. 

Sourav Bhaduri supervised the research, contributed to study design and interpretation of results, and critically revised the manuscript.
% =======================================================
\section*{Competing interests}
% =======================================================

The author declares no competing interests.

% =======================================================
\section*{Acknowledgements}
% =======================================================

The Framingham Heart Study is conducted and supported by the
National Heart, Lung, and Blood Institute (NHLBI) in collaboration
with Boston University.
This manuscript was not prepared in collaboration with investigators
of the Framingham Heart Study and does not necessarily reflect the
opinions or views of the Framingham Heart Study, Boston University,
or NHLBI.

\FloatBarrier

% =======================================================
% BIBLIOGRAPHY
% =======================================================

\end{document}